\documentclass{article}

\title{Continuous opinion dynamics of multidimensional allocation problems under bounded confidence:\\ More dimensions lead to better chances for consensus}

\author{Jan Lorenz \\ Universität Bremen, Fachbereich Mathmatik und Informatik\\ D-28359 Bremen\\[3pt] http://www.janlo.de, post@janlo.de\\[6pt] }

\date{17 November 2005, presented at Conference ''Connectionist Approaches in Economics and Management - ASCEG'', Aix-en-Provence, France}

\usepackage{array}
\usepackage{theorem}
\usepackage{latexsym}
\usepackage{graphicx}
\usepackage{epsfig}
\usepackage{amsfonts}
\usepackage{amsmath}
\usepackage[abbr,dcucite]{harvard}

\newcommand{\R}{\mathbb{R}}
\newcommand{\N}{\mathbb{N}}

\newcommand{\ba}[1]{\begin{array}{#1}}
\newcommand{\ea}{\end{array}}

\newcommand{\eps}{\varepsilon}
\newcommand{\n}{\underline{n}}
\newcommand{\m}{\underline{m}}
\newcommand{\Vol}{\mathrm{Vol}}

\newtheorem{defn}{Definition}

\begin{document}

\maketitle

\begin{abstract}
We study multidimensional continuous opinion dynamics, where
opinions are nonnegative vectors which components sum up to one.
Examples of such opinions are budgets or other allocation vectors
which display a distribution of a fixed amount of ressource to $n$
projects.\\
We use the opinion dynamics models of Deffuant-Weisbuch
and Hegselmann-Krause, which both extend naturally to more
dimensional opinions. They both rely on bounded confidence of the
agents and differ in their communication regime. We show detailed
simulation results regarding $n=2,\dots,8$ and the bound of
confidence $\eps$. Number, location and size of opinion clusters
in the stabilized opinion profiles are of interest.\\
Known differences of both models repeat under higher opinion
dimensions: Higher number of clusters and more minor clusters in
the Deffuant-Weisbuch model, meta-stable states in the
Hegselmann-Krause model. But surprisingly, higher dimensions lead
to better chances for a vast majority consensus even for lower
bounds of confidence. On the other hand, the number of minority
clusters rises with $n$, too.
\end{abstract}


\section{Introduction}
Consider a group of $m$ agents, each having an opinion about a
certain issue. Consider that each opinion is a $n$-dimensional
nonnegative vector of real numbers, which components sum up to
one, e.g. the allocation of a fixed amount of money to $n$
projects or the probabilities one admits to $n$ propositions. The
example we want to stress here is that each opinion is a budget
plan proposal of an agent. So each agent has a proposal how to
distribute a fixed amount of money to $n$ departments. Such
opinions should be held by members of a parliament which have to
discuss on a states budget or by citizens of a town which have to
work in a participatory budgeting process on the cities budget
plan Participatory budgeting has been invented in 1989 in the
Brazilian city of Porto Alegre and is now tested in several local
communities all around the world \cite{Santos1998,Lorenz2005c}.
One argument for participatory budgeting is, that it fosters
social consensus. This simulation gives some evidence that raising
the number of decision parameter may indeed foster the evolution
of consensus in the discussion.

The group of agents (politicians or citizens) is to find an
agreement about the allocation. We suppose that each agent is
willing to revise his allocation vector by taking the opinion
vectors (budget plan proposals) of other competent agents into
consideration. A competent agent in the view of one agent should
be an agent with an opinion which is in a measurable way\footnote{
the Euclidean distance in this study} not more than $\eps$ away
from his own opinion. $\eps$ is called the \emph{bound of
confidence}. This process of repeated discussing and revising of
opinions is called \emph{continuous\footnote{'Continuous' refers
to the type of opinions not to the time.} opinion dynamics under
bounded confidence}.

Continuous opinion dynamics of one-dimensional opinions under
bounded confidence has been studied recently under different
communication regimes. In the model of Deffuant-Weisbuch (DW)
\cite{Deffuant2000,Weisbuch2002} two random agents meet in each
time step and compromise at the arithmetic mean\footnote{With $\mu
= 0.5$.} of their two opinions if their difference in opinion is
beneath the bound of confidence $\eps$. In the model of
Hegelmann-Krause (HK) \cite{Krause2000,Hegselmann2002} all agents
revise their opinions at the same time. Obviously, both models
differ only in their communication regime. They have been
generalized into one model by Urbig and Lorenz \cite{Urbig2004}.
To some extend each model represents the extrem point of possible
communication regimes. While the HK model relies on the full
knowledge of every opinion (e.g. in a general meeting where
everyone communicates his opinion), the DW model relies totally on
'gossip' in random pairwise encounters of agents. We will use both
models in our analysis, suggesting that real communication
behavior lies somewhere in between. Thus, we hope that these two
models serve as extreme points regarding communication structure.

Both models have been analysed roughly for heterogeneous bounds of
confidence \cite{Weisbuch2002,Lorenz2003} and on different
network topologies \cite{Amblard2004,Fortunato2005}. The DW model
has been extended e.\,g.\ to extremism \cite{Deffuant2002} and to
limited verbalisation capabilities of the agents \cite{Urbig2003}.
In \cite{Hegselmann2004} there are several extensions for the HK
model. In contrast to this mass of extensions, studies of the
multidimensional case are rare, although both models extend
naturally to real-valued vectors. The only mathematical thing to
define new is the distance measure for vectors.

A short review of what exists with more dimensions. The vector
opinions in \cite{Weisbuch2002} are bit-strings. The same holds
for the Axelrod's famous model of cultural dissemination
\cite{Axelrod1997,Laguna2003}\footnote{In Axelrod's model a higher
number of cultural features fosters the evolution of cultural
homogeneity. Similarities to our results are discussed in section
\ref{sec:conclusion}.}. Thus, these models have no continuous
opinion space and regard different opinion issues.
\cite{Jager2005,Urbig2005} made interesting studies with two
dimensional opinions but focussing on specific interplays between
two opinion dimensions.

The naturally extended models have been studied in
\cite{Lorenz2003b,Fortunato2005b} but not under the budgeting
restriction of sum-one vectors. Probably, most simulators hesitate
to simulate the multidimensional case, because modelling and
systematic characterisation of the parameter space gets
indistinct. The only expedient onedimensional opinion spaces are
intervals (e.g. $[0,1]$). But in higher dimensions every convex
set seems expedient. The same holds for the distance measure
regarding the bound of confidence.

We decided to use the nonnegative sum-one vectors as opinion space
(known as the unit simplex) for three reasons. First, we have a an
important application with the issue of budgeting. Second, the
unit simplex is the most simple polygon in higher dimension. For a
fixed dimension, it has the lowest possible number of edges and
faces. While opinion dynamics is known to be driven from the
borders of the opinion space, we regard the unit simplex as the
most simple case under maximal number of dimension, a structural
argument.

And third we can derive the use of the arithmetic mean in
multidimensional opinion dynamics from the axiomatisation of
allocation aggregations in \cite{Lehrer1981}(page 112): If there
are at least three departments and if the aggregation function
assigns the allocation to each department purely as a function of
the allocations to that department by the agents (irrelevance of
alternatives), and respects their agreement in assigning a
department the amount zero (zero unanimity), then the aggregation
function is a weighted arithmetic mean.

In our models the revision an agent does with his opinion is an
allocation aggregation of the opinions of others. Thus, if we
extend our models to more than two dimensions and restrict them to
allocation problems, we got a rationale to use the arithmetic mean
by the axiomatisation (which we had not in the one-dimensional
case!). (See \cite{Hegselmann2004} for other aggregation methods
in the onedimensional case).

In the next section we will define the models (extended to
multidimensionality) and give a review and comparison about the
facts we know about the two onedimensional models. Section
\ref{sec:sim} shows the simulation results and gives some
explanations by analysis of the geometry of the opinion space.
Section \ref{sec:conclusion} gives conclusions.

One disclaimer in advance: The simulation of these models is not
to forecast real opinion dynamics, but to give qualitative hints
about underlying inherent dynamics of opinion formation.

\section{The models and what we know}

\subparagraph{General definitions}
Let $\m := \{1,\dots,m\}$ be the set of agents and $\n :=
\{1,\dots,n\}$ the set of opinion dimensions. We define the
\emph{$k$-Simplex} $\Delta^k := \{x \in \R^{k+1} \,|\, x\geq 0,
\sum_{i=1}^{k+1}x_i = 1\}$ which is our \emph{opinion space}.
Attention, the $k$-Simplex is a subset of $\R^{k+1}$, this is a
mathematical convention because the $k$-Simplex has affine
dimension $k$. This is clear due to the fact that every $k$
dimensions automatically define the last dimension due to
$x_{k+1}=1-(x_1+\dots +x_k)$. Thus, an opinion space with $n$
opinion dimensions is $\Delta^{n-1}$.

We will call $x^i\in\Delta^{n-1}$ the opinion of agent $i\in\m$.
We call the vector of all opinion vectors $X(t) \in
(\Delta^{n-1})^m$ the \emph{opinion profile} at time step $t$.
Thus $x^j_i(t)$ is the opinion of agent $i$ about dimension $j$ at
time step $t$.

For the further development of the formalism it makes sense to
think of the vector of vectors $X$ as a matrix $m\times n$, where
each row represents an agents opinion. (The opinion is a row
vector here.)

Both models need a \emph{bound of confidence} $\eps \in \R_{>0}$
and a measure of distance of two (multidimensional) opinions $x^1,
x^2 \in \Delta^n$. In this paper we will use the Euclidean
distance $||x^1-x^2||$ (derived from the Euclidean norm). Thus,
$||\cdot||$ stands for the Euclidean norm in the respective
multidimensional space throughout the paper. Surely, other
distance measures are possible and interesting for further
research.

We will define for both models the \emph{process of continuous
opinion dynamics} as a sequence of opinion profiles.

\begin{defn}[Deffuant-Weisbuch Model]\footnote{This is a natural multidimensional extension of the basic
version of the model in \cite{Weisbuch2002} with $\mu=0.5$.} Given
an initial profile $X(0) \in (\Delta^{n-1})^m$ and a bound of
confidence $\eps\in\R_{>0}$ we define the \emph{Deffuant-Weisbuch
process of opinion dynamics} as the random process
$(X(t))_{t\in\N_0}$ that chooses in each time step $t\in\N_0$ two
random\footnote{With 'random' we mean 'random and equally
distributed in the respective space'.} agents $i,j \in \m$ which
perform the action
\begin{eqnarray*}
x^i(t+1) &=& \left\{
\begin{array}{ll}
    \frac{1}{2}(x^i(t)+x^j(t)) & \hbox{if $||x^i(t)-x^j(t)||\leq\eps$} \\
    x^i(t) & \hbox{otherwise.} \\
\end{array}
\right.\\
\end{eqnarray*}
The same for $x^j(t+1)$ with $i$ and $j$ interchanged.
\end{defn}

\begin{defn}[Hegselmann-Krause Model]\footnote{This a natural multidimensional extension of the model in \cite{Hegselmann2002}.} Given a bound of
confidence $\eps\in\R_{>0}$ we define for an opinion profile $X\in
(\Delta^{n-1})^m$ the \emph{confidence matrix} $A(X,\eps)$ as
\[
 a_{ij}(X,\eps) :=
 \left\{ \begin{array}{cl}
   \frac{1}{\#I(i,X)} \quad & \textrm{if } j\in I(i,X)   \\
   0 & \textrm{otherwise,}
\end{array} \right. \\
\]
with $I(i,X) := \{j \in \n \,|\, ||x^i - x^j|| \leq \eps \}$.
("$\#$" stands for the number of elements.)

Given an initial opinion profile $X(0) \in \R^n$, we define the
\emph{Hegselmann-Krause process of opinion dynamics} as a sequence
of opinion profiles $(X(t))_{t\in\N_0}$ recursively defined
through
\[X(t+1) = A(X(t),\eps) X(t)\]
\end{defn}

It has been proved analytically in \cite{Lorenz2005} that both
processes converge to a stabilized opinion profile for every
initial condition. Thus each process of continuous opinion
dynamics converges to a stabilised opinion profile.\footnote{The
first proof of stabilisation has been done in \cite{Dittmer2001},
but it holds only in the one-dimensional case.}

One wants to know now how the stabilised profile looks like. But
this depends heavily on the initial profile and for the DW process
also on the choice of agents who meet. Thus, simulation has to
play its role. Actually, both models have been studied more by
simulation then analytically.

\subparagraph{Facts about the one-dimensional case}

In nearly every study one takes initial profiles with random and
equally distributed opinions between zero and one.

To circumvent the specific properties of one random setting one
studies a set of several random initial profiles and computes all
stabilised profiles. This set of stabilised profiles is analysed
for clusters (e.g. opinion blocks or, political parties). The
parameters of most interest are the number, the location and the
size of clusters.

\begin{figure}[htbp]
\includegraphics[width=\columnwidth]{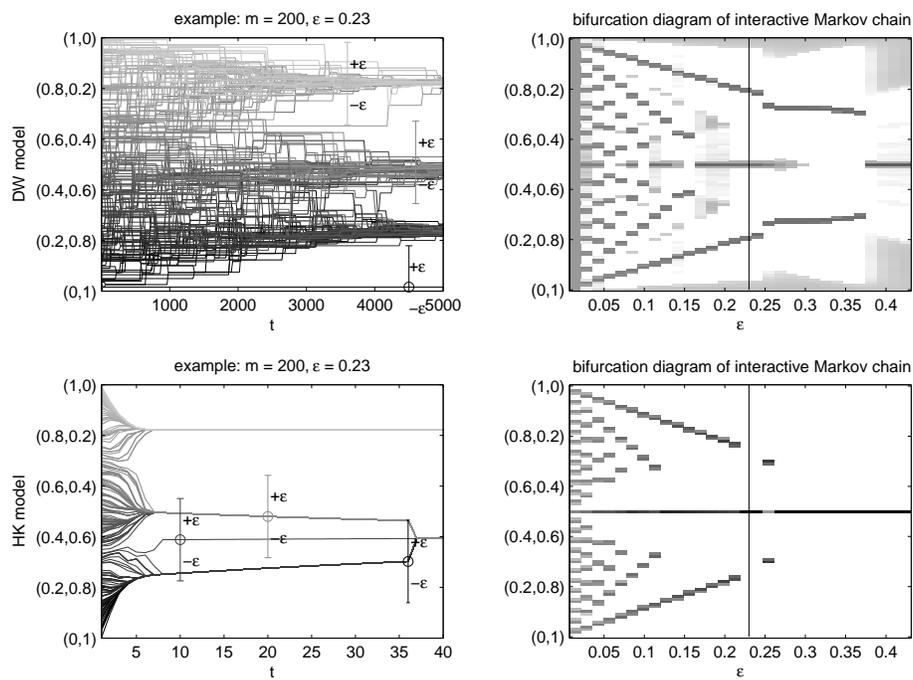}
\caption{Overview about the one-dimensional case (two opinion
dimensions in the allocation case) by example processes and
bifurcation diagrams derived for interactive Markov chains.}
\label{fig1}
\end{figure}

In Figure \ref{fig1} we summarize what we know about the two
models.

\subparagraph{Description of figure \ref{fig1}} We give an example
process with $m=200$ agents and the bifurcation diagram dervied
from the governing interactive Markov chain for each model. (See
\cite{Lorenz2005a,Lorenz2006} for details about the interactive
Markov chain.) We set the opinion space to the line
$[(0,1),(1,0)]$ which is equal to $\Delta^1$ and has a Euclidean
length $\sqrt{2}$ to get a good comparison with further
figures.\footnote{For comparison of the results with
\cite{Weisbuch2002,Hegselmann2002} we have to respect that the
length of the opinion space there is one. We can compare our
$\eps$ with $\sqrt{2}\eps$ in \cite{Weisbuch2002,Hegselmann2002}.}

A quick description how the dynamic works: Agents will move to
opinion regions with a high density of agents. Thus, in our case
of uniform distribution the dynamic will start at the border of
the opinion space with agents moving towards the center, creating
a higher density there and thus attracting other agents even from
the center.

We plot three confidence intervals around selected opinions in
each of the two example processes to give an impression how far
the confidence of one agent reaches. The vertical line in the
bifurcation diagrams represents the $\eps$-value $(0.23)$ of the
example.

The bifurcation diagrams we display are not as usual in
bifurcation theory but 'reverse'. We have to read them from right
to left to say that clusters bifurcate (or split) into more
clusters. The $\eps$-value where this happens is called a
\emph{bifurcation point}. The diagrams are very familiar with the
bifurcation diagram in \cite{Ben-Naim2003} (which is not reverse).
There is also a relationship to the figures in
\cite{Hegselmann2002}. In application we often look for the
\emph{consensus brink}\footnote{The term goes back to
\cite{Hegselmann2004a}.} which is the bifurcation point form
consensus to polarization. With this term in mind it is more
natural to think of rising $\eps$.

One observation in the bifurcation diagrams of figure \ref{fig1}
is that the consensus brink is significantly lower in the HK model
as in the WD model. The following simulation study will show that
the consensus brink can be lowered in both models by raising the
number of dimensions under the restriction to allocation problems.

\section{The impact of multidimensionality}\label{sec:sim}

\subparagraph{Simulation setup} The basis for our simulation
results are 200 random opinion profiles out of $(\Delta^1)^{200},
\dots , (\Delta^7)^{200}$, so $n=2,\dots,8$. Choosing a random
opinion uniformly distributed in a simplex $\Delta^k$ is not
totally trivial. To ensure a uniform distribution and to avoid
distortions by normalisation of random vectors we take a random
vector out of $[0,1]^k$ and compute one more entry as 1 minus the
sum of all other entries, but only if the sum of all former
components is less or equal than 1. If we fail, we try it
again.\footnote{Obviously, the probability for success is getting
rapidly low for rising $k$. For that reason we had to restrict us
at $n=8$ for computation time reasons.}

Further on, we explore the parameter space $\eps =
0.15,\stackrel{+0.01}{\dots},0.42$. Thus, we ran 200 simulations
runs for each pair $(n,\eps)$. Additionally, we took for $n=3$ a
set of 4000 initial profiles to compute the stabilized profiles
for selected values of $\eps$ to start the exploration of more
opinion dimensions.

\begin{figure}[htbp]
\includegraphics[width=\columnwidth]{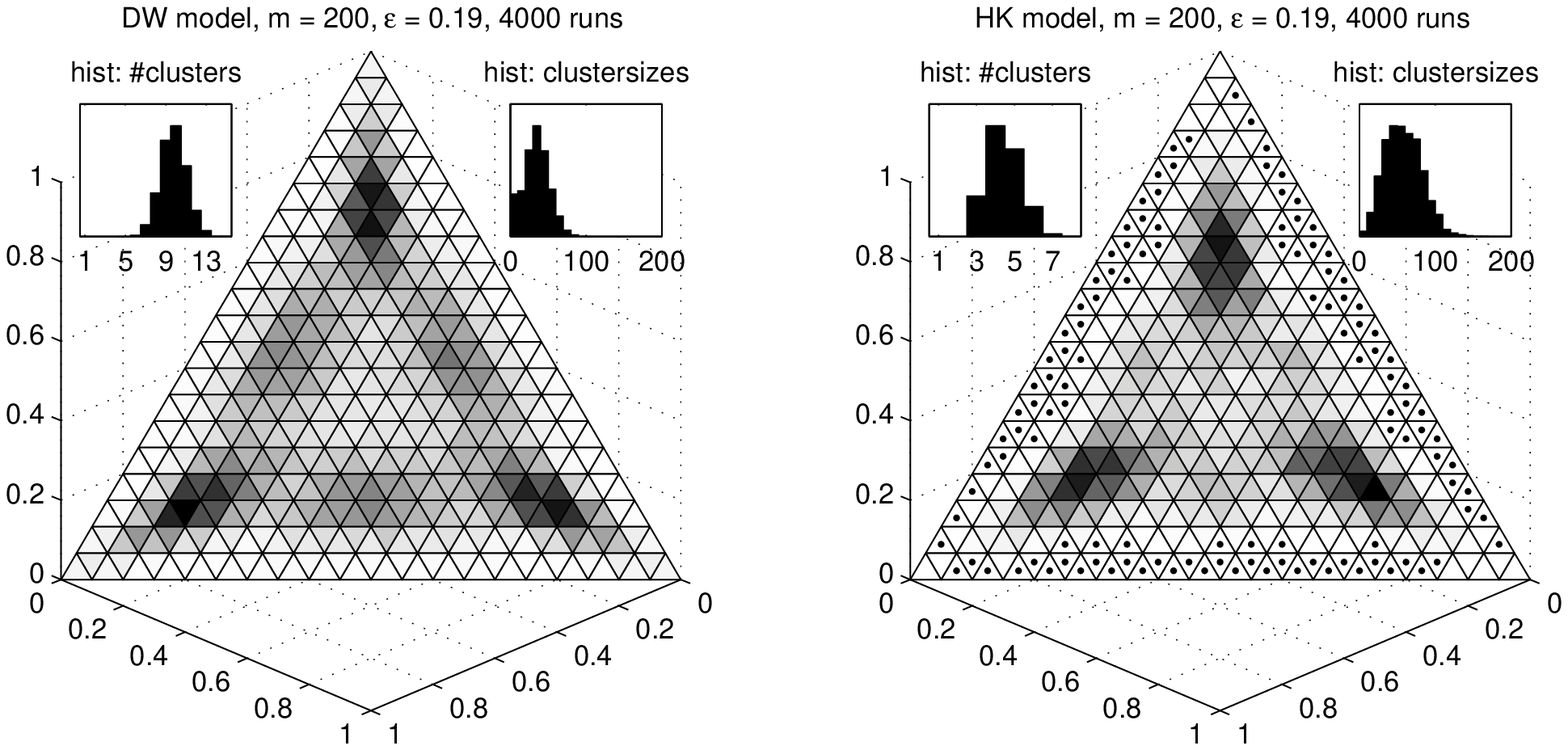}
\includegraphics[width=\columnwidth]{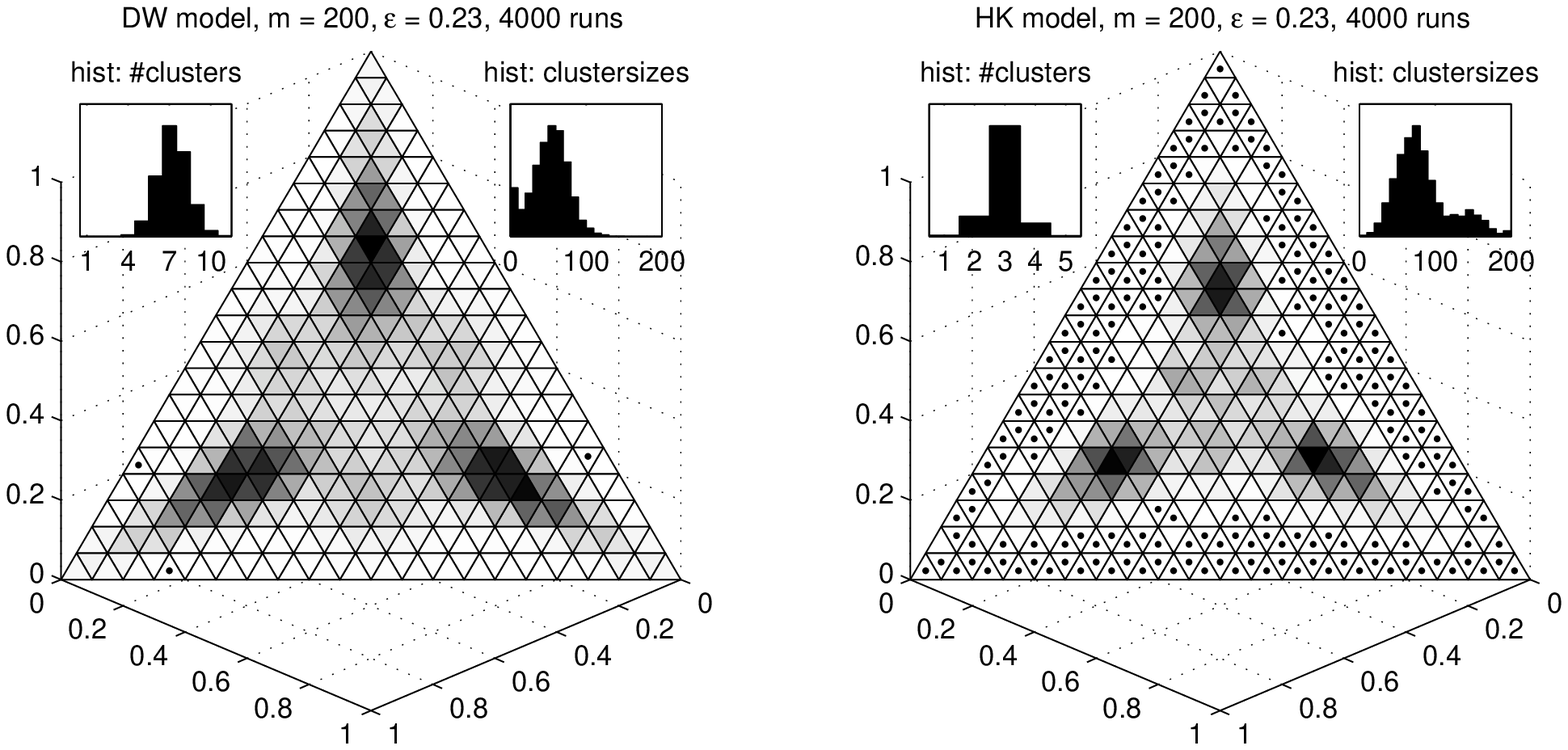}
\includegraphics[width=\columnwidth]{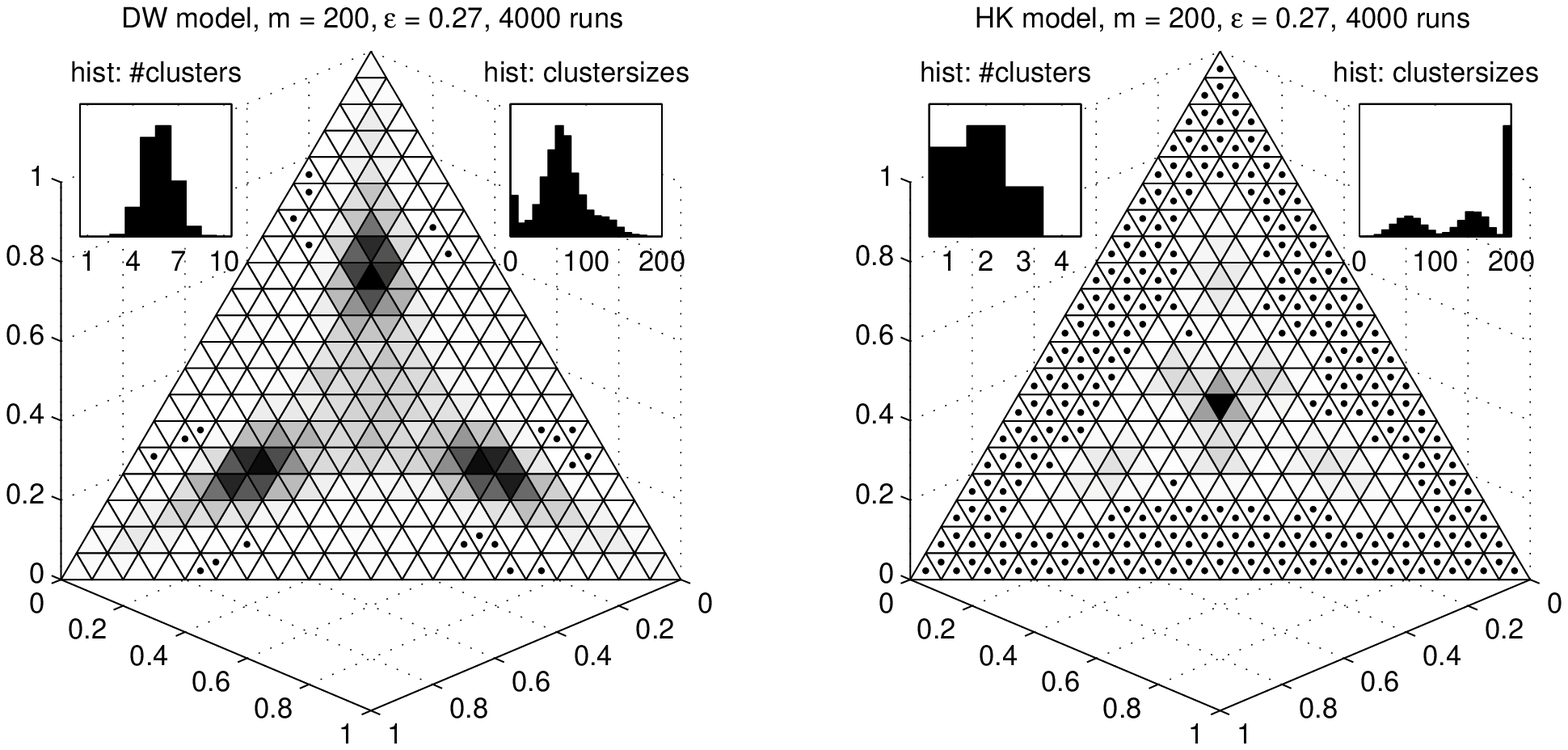}
\caption{Overview about location and sizes of clusters with 3
opinion dimensions for selected $\eps$-values.}\label{fig2}
\end{figure}

\begin{figure}[htbp]
\includegraphics[width=\columnwidth]{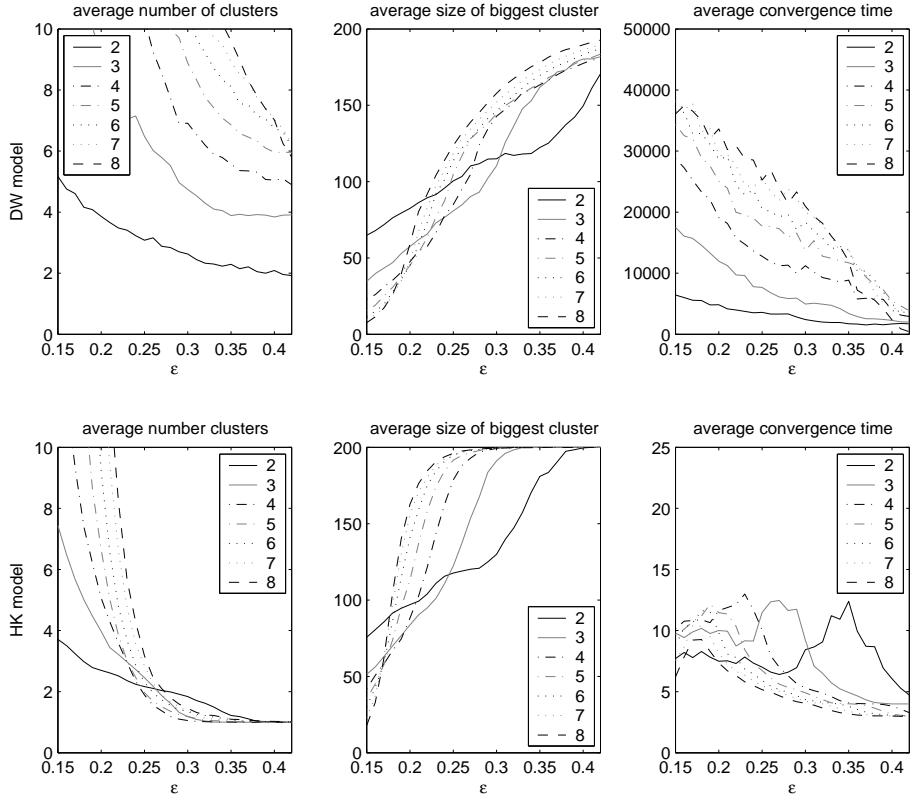}
\caption{Average number of clusters, size of the biggest cluster
and convergence time for $n=2,\dots,8$.} \label{fig3}
\end{figure}

\begin{figure}[htbp]
\includegraphics[width=\columnwidth]{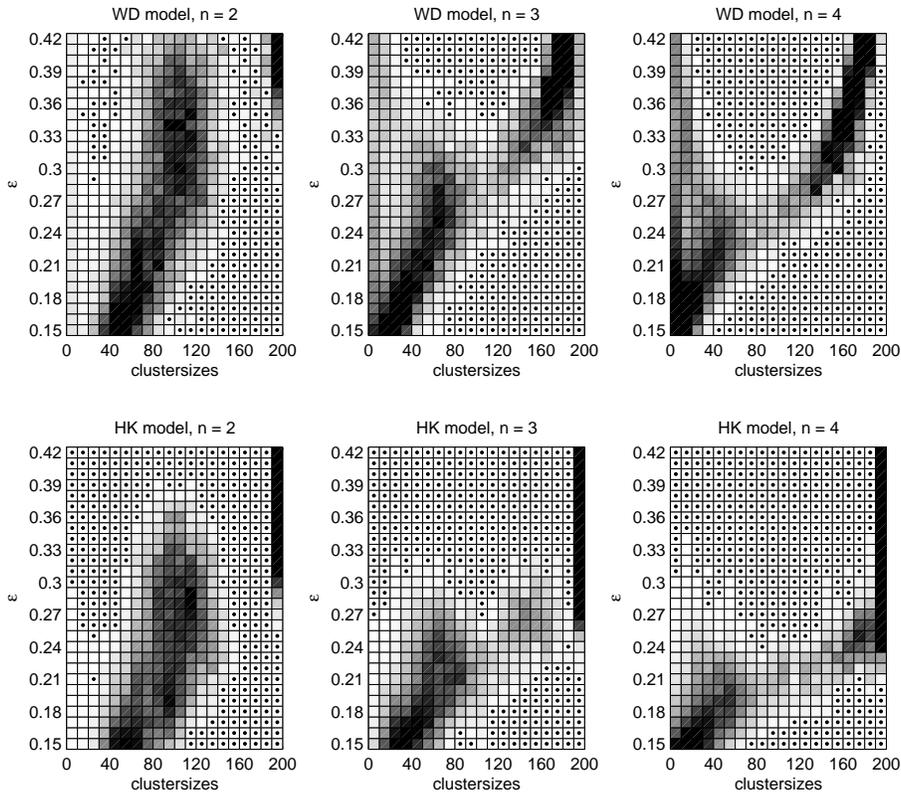}
\caption{Histograms of the sizes of clusters for
$n=2,3,4$.}\label{fig4}
\end{figure}

\begin{figure}[htbp]
\includegraphics[width=\columnwidth]{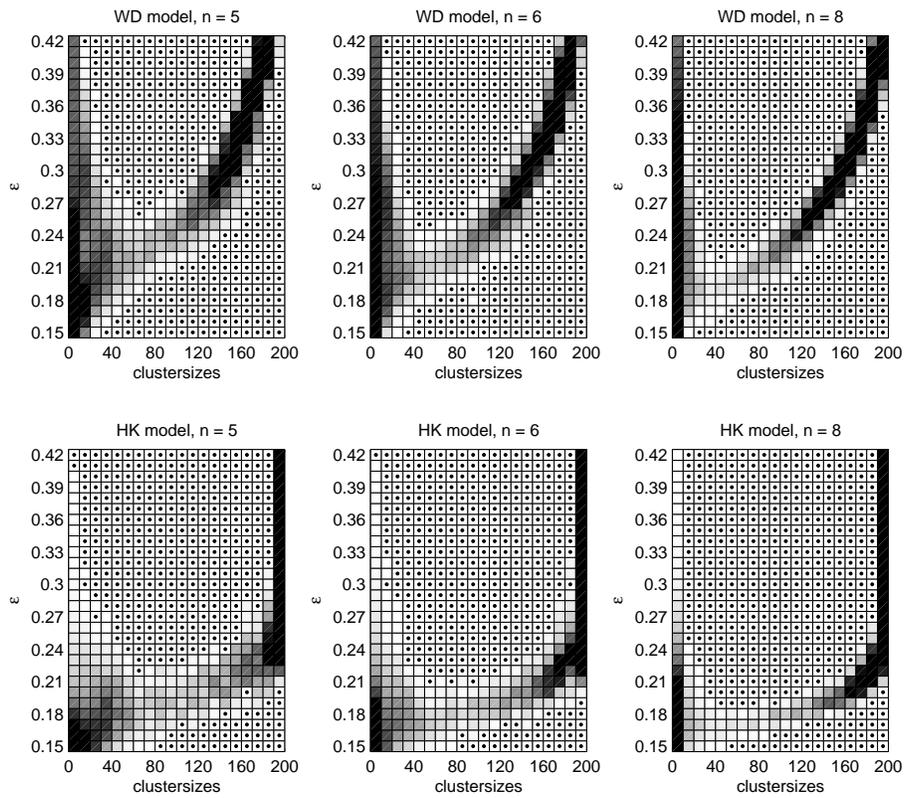}
\caption{Histograms of the sizes of clusters for
$n=5,6,8$.}\label{fig5}
\end{figure}

Figures \ref{fig2}, \ref{fig3}, \ref{fig4} and \ref{fig5} show
visualisations of relevant properties of the set of stabilised
profiles. In the following we will deliver descriptions of the
figures, a summary about the dynamics with three departments
($n=3$) and a summary about the impact the number of departments
$n$.

\subparagraph{Description of Figure \ref{fig2}} We see the opinion
space $\Delta^2$ divided into several sub-triangles. The
gray-scale of one triangle stands for the number of opinions of
all 4000 stabilised profiles which are in that region (black is
high, white low, and a spot is zero). The gray scale is only
relative in each figure. It shows attractive regions for clusters.
Additionally, we got two histograms in each subfigure. The
histogram \#clusters shows how often a stabilized profile got a
specific number of clusters in all 4000 runs. It is important to
notice that we count each cluster even isolated outliers. In the
histogram about clustersizes we group all agents by the size of
the cluster where they belong to. The bins are $1-10, 11-20,\dots,
191-200$.

\subparagraph{Summary about dynamics with three departments
($n=3$)} Most of the results can be derived from figure
\ref{fig2}.

\begin{enumerate} \item The characteristic polarization
into two big clusters for a great $\eps$-interval in $\Delta^1$
extends to a characteristic polarization into three big clusters
in $\Delta^2$. Each cluster represents a budget plan proposal
like: "The biggest part for one department (around 70\%) and the
rest equally for the two others."

\item A polarization into two opinion clusters is also possible.
It occurs by the union of two of the three characteristic
clusters, but we can not predict which two clusters unite. In this
situation we have two third of the agents saying about
"40\%/40\%/20\%" and the other third "15\%/15\%/70\%".

\item Higher numbers of big clusters and more chances for
minorities to survive in the DW model as reported in the one
dimensional model occur also in $\Delta^2$.
\end{enumerate}

\subparagraph{Description of Figure \ref{fig3}} We collect some
aggregated data about the 200 stabilised profiles for each pair
$(n,\eps)$ and the two models. The left hand figure column shows
the average number of clusters (arithmetic mean over all 200
runs). We count every cluster, even if it is only an isolated
agent. Thus, another interesting parameter is the average size of
the biggest cluster, which gives an impression how many agents
have found a common agreement. This is shown in the central column
figure. The right hand column shows average time to reach
stabilisation. Observing stabilisation in the HK model is easy by
checking if something has changed in one step. This does not work
in the DW model. We use this: Every 2000 time steps we check the
opinion profile for clusters (a set of directly or indirectly
connected agents), if in each cluster the maximal distance of
agents is below $\eps$, we stop the time. Fortunately, we can
compute the final location of the cluster by building the
arithmetic mean (see \cite{Urbig2004} for details). Thus the
convergence times in the figure is counted only to the time when
clusters have built which can not split anymore, and the time we
measure may be at least 1999 time steps to long.

\subparagraph{Description of Figure \ref{fig4} and \ref{fig5}} In
each row we got the histogram about the clustersizes. We group all
agents by the size of the cluster where they belong to. The bins
are $1-10, 11-20,\dots, 191-200$. This is the same as the
histogram in figure \ref{fig2} at the right hand side of each
subfigure. The gray-scale of each histogram is only relative in
each row. Black is a high number of agents in clusters of that
size. White is a low number and a dot is no clusters of that size.
We omit the subfigure for $n=7$ it has no additional information.

\subparagraph{The impact of the number of departments ($n$)} This
paragraph summarizes the key results of this paper. The
conclusions can be derived from the figures \ref{fig3}, \ref{fig4}
and \ref{fig5}.
\begin{enumerate}
\item Raising the numbers of departments leads to more minor
clusters not only in the DW model, but also in the HK model.

\item If we regard the existence of a cluster with a vast majority
of agents (e.g. more than 150, so 75\%) as a
\emph{majority-consensus} we can say that the majority-consensus
brink is sinking with rising $n$. But sinking slows down.

\item The $\eps$-interval between majority consensus and total
plurality of opinions is getting shorter with rising $n$.

\item The convergence times in the HK model give a hint that we
may reach meta-stable states with long convergence times in the
$\eps$-interval of majority-consensus close to the
majority-consensus brink.
\end{enumerate}

\subparagraph{About the multidimensional opinion space}
In this paragraph we give some geometrical facts about the opinion
space $\Delta^k$ which shed some light on the lowering of the
majority consensus brink with rising $n$.

According to \cite{Buchholz1992} the $k$-dimensional volume of
$\Delta^k$ is
\[\Vol\Delta^k = \frac{\sqrt{k+1}}{k!} = \frac{\sqrt{k+1}}{\Gamma(k+1)}.\]
(The $\Gamma$-function is an extension of $n!$ to real numbers. It
holds $\Gamma(n+1)=n!$.) The $n$-dimensional volume of the ball
$B_\eps^n$(regarding the Euclidean norm) with radius $\eps$ (our
area of confidence) is
\[\Vol B_\eps^n = \eps^n \frac{\pi^{n/2}}{\Gamma(\frac{n}{2}+1)}.\]
For that reason it holds that for each $\eps$ there is an $n$ such
that $\Vol B_\eps^n>\Vol\Delta^k$ which means that the area of
confidence is bigger than the whole opinion space. To give some
numerical values: $\eps=0.23$ it is for $n\geq 10$, for
$\eps=0.15$ it is for $n\geq 21$. That explains why we get that
strong impact on majority-consensus with rising $n$.

On the other hand the Euclidean distance from one extreme opinion,
e.g. $(1,0,\dots,0)$, to the barycenter of the opinion space
$\frac{1}{n}(1,\dots,1)$ (which is the attractive point for
consensus) is $\sqrt{(1-\frac{1}{n})^2 + (n-1)\frac{1}{n^2}}$ and
is thus converging to 1 for $n\to \infty$ (which means that it
comes infinitely close to the barycenter of each $n-1$-dimensional
face of $\Delta^n$). This explains the growing probability for
extremists to get cut of the others during the process.

Colloquial: The opinion space of allocation vectors transforms
with growing number of departments to a space with greater
compromising opportunities in the center but growing hiding-places
for extremists.

Actually, in our random profiles of $\Delta^7$ some extreme agents
were disconnected from the other agents from the very beginning
(for the lower $\eps$-values).

Thus, the geometry of the simplex opinion space is a major source
of the phenomena described above. Performing the same analysis
with e.g. a cubic opinion space will probably not show such
drastic effects when raising the dimension of the cube (for some
hints see \cite{Lorenz2003b}.

\section{Conclusion}\label{sec:conclusion}
We come back to the issue of social consensus. From our simulation
results one may conclude that a public debate and decision about a
continuous issue may come to a social majority consensus more
likely if we give not only two opposite possibilities but more to
allocate the resources between. This holds under a gossip like
communication (DW) as well as under a meeting driven communication
(HK).

Main assumptions of the models were bounded confidence of all
agents, arithmetic averaging as opinion aggregation and a uniform
distribution of opinions at the beginning. This is obviously not
the case in each real situations when budget discussion begins.
Long time ideologies, hard positions regarding certain departments
and pre-clustering by political parties and lobby groups certainly
play important roles. Probably, most of these issues do not have a
good impact on the chances of social consensus. Simplifying the
opinion space by projecting it to a lower dimensional (e.g.
twodimensional) opinion space as it is often done by mass media or
populists, may, due to our results, significantly lower chances of
reaching a social consensus.

A main impact factor is the structure of the opinion space itself.
Thus, the restriction to a fixed amount of money to distribute
(without permitting negative amounts) has a surprisingly strong
force to attract agents to the consensual center. Colloquial: The
fact that one can only suggest a higher budget for one department
if one lowers the budgets of other departments produces an opinion
space which is more appropriate to foster social consensus.

Nevertheless, the impact of fostering social majority consensus by
raising the number of dimensions is sinking if we have already a
multidimensional opinion space. Raising from 7 to 8 dimensions has
not the same impact as raising from 2 to 3. The drawback of
raising the number of dimensions is that we open new space for
extremists in each corner, which gets more and more loosely
connected to the center. Thus, the social majority consensus is
getting more and more a slim majority consensus, especially if we
rely only on gossip dynamics.

\end{document}